\documentclass[prd, twocolumn, nofootinbib]{revtex4}
\usepackage[dvips]{graphics}
\usepackage{amsmath, amssymb}
\newcommand{\z}{\psi}
\newcommand{\ddt}{\frac{\partial}{\partial t}}
\newcommand{\ddr}{\frac{\partial}{\partial r}}
\newcommand{\lie}{\pounds}
\renewcommand{\leq}{\leqslant}
\renewcommand{\geq}{\geqslant}
\newcommand{\scri}{\mathcal{I}}
\newcommand{\R}{\mathbb{R}}
\newcommand{\g}{g}
\newcommand{\G}{{}^{(5)}g}
\newcommand{\tg}{\tilde{g}}
\newcommand{\ricci}{R}
\newcommand{\Ricci}{{}^{(5)}R}
\newcommand{\fourRicci}{R_\mathrm{S}}
\newcommand{\tricci}{\tilde{R}}
\newcommand{\eins}{G}
\newcommand{\Eins}{{}^{(5)}G}
\newcommand{\teins}{\tilde{G}}
\newcommand{\stress}{T}
\newcommand{\Stress}{{}^{(5)}T}
\newcommand{\SStress}{{}^{(4+n)}T}

\newcommand{\tD}{\tilde{D}}
\newcommand{\refsec}{section}

\begin{document}
\title{The Collapse of Large Extra Dimensions}
\author{James Geddes}
\affiliation{Enrico Fermi Institute and Department of Physics \\
University of Chicago \\
5640 S.\ Ellis Avenue, Chicago, IL 60637, USA}
\begin{abstract}
In models of spacetime that are the product of a four-dimensional
spacetime with an ``extra'' dimension, there is the possibility that
the extra dimension will collapse to zero size, forming a
singularity. We ask whether this collapse is likely to destroy the
spacetime. We argue, by an appeal to the four-dimensional cosmic
censorship conjecture, that---at least in the case when the extra
dimension is homogeneous---such a collapse will lead to a singularity
hidden within a black string. We also construct explicit initial data
for a spacetime in which such a collapse is guaranteed to occur and
show how the formation of a naked singularity is likely avoided.
\end{abstract}
\pacs{04.20.Dw}
\maketitle

\section{Introduction}
The idea that we live in a universe with more than the four dimensions
we observe has been around for some time. Models of the universe with
five or more dimensions, originally proposed by Kaluza and Klein
\cite{kaluza,klein} as an attempt to unify electromagnetism and
general relativity, have been commonplace in string theory for many
years. In such theories the extra dimensions typically have a ``size''
comparable to the Planck length and thus remain unseen since
experiments that would reveal their presence require
as-yet-unattainable energies. Furthermore, questions about the
evolution and stability of the extra dimensions have been largely
ignored since at this scale quantum gravity effects are presumably
important and it is difficult to extract predictions from any current
candidate theory of quantum gravity.

Recently, however, there has been a great deal of interest in models
wherein the size of the extra dimensions is much larger than the
Planck length~\cite{ah, anton, rubakovreview}. Current experimental
results involving tests of the inverse square law (see, e.g., Hoyle
\textit{et al.}~\cite{experiment}) do not rule out extra
dimensions even as large as a tenth of a millimeter.\footnote{In order
that the extra dimensions remain unobserved, one imagines that the
standard model fields are confined to a four-dimensional submanifold,
known as the ``brane,'' which comprises the observable universe. In
what follows we ignore the existence of the brane. There have been
some attempts to model the brane in a theoretically reasonable way as
a distributional stress-energy~\cite{rs,rs1}, albeit with a
non-compact extra dimension, but we shall assume that the
stress-energy of the brane can be ignored in comparison to the
stress-energy in the full spacetime.}  It is now important to consider
the evolution of the extra dimensions since the observed strength of
the gravitational force is directly dependent on the size of the extra
dimensions.\footnote{Indeed, it is for this reason that these models
were proposed in the first place: by fixing the gravitational field
strength appropriately, one can arrange for the actual Planck energy
to be comparable to the electroweak scale yet explain the size of the
observed Planck energy by this weakening of the observed gravitational
field strength on the brane. It was suggested that one thereby
explains the surprising weakness of gravity compared to the other
forces, although to some extent the problem has merely been
transferred to explaining the size of the extra dimensions.}
Furthermore, since the curvature of spacetime is now much larger than
Planckian scales it ought to be possible to study the evolution of
such spacetimes within the framework of classical general relativity.

As an example, consider a spacetime whose manifold is the product of
four-dimensional Minkowski spacetime with a single extra dimension of
topology $S^1$ and whose metric is
\[
ds^2 = \eta_{\mu\nu}\,dx^\mu\, dx^\nu+b(t)^2d\z^2.
\]
Here $\eta_{\mu\nu}$ is the metric of Minkowski spacetime, $x^\mu$ are
the coordinates in ``usual'' dimensions, $\z$ is the coordinate in
the fifth dimension, and $b(t)$ is the scale of the extra
dimension. It is clear that that this metric is a solution to
Einstein's equation when $b(t)$ is constant, say $b(t)=b_0$, since the
spacetime is then flat. However, it is easy to check that a solution
is also obtained by setting $b(t)=b_0+\alpha t$, with $\alpha$ a
constant. If $\alpha$ is negative, then clearly the extra dimensions
will collapse to zero size---and the whole spacetime will become
singular---in finite time. Although this model is rather unrealistic
in that the scale factor of the extra dimension is the same, and
evolving in the same manner, throughout the entire space, we shall
show in \refsec~\ref{chap:collapsemodel} that it is possible to
construct more realistic examples in which the collapse happens
locally (i.e., within some compact spatial region) and is guaranteed
to produce a singularity. 

There do exist models in which the size of the extra dimensions is
stabilized, at least under small perturbations, by the addition of
suitable matter~\cite{gw, rs, guenther97, classicalstab, stab}. However, it is
still not clear whether any of these models would describe our
universe if extra-dimensional models were taken seriously. Thus one
must be concerned about the possibility of singularity formation in
the fashion described above and the nature of the singularity so
formed. It would be disastrous, for example, if a singularity, once
formed, were to propagate outwards from its origin, destroying the
spacetime.

Nonetheless, we shall argue that, under reasonable assumptions, a
space that is the metric product of a three-dimensional space and an
homogeneous, one-dimensional manifold, in which the scale-factor of
the extra dimension is collapsing to zero in some region, will evolve
to a ``black string;'' that is, a spacetime that is the metric product
of a four-dimensional black-hole spacetime with the extra-dimensional
manifold. That is, even if a singularity is formed by
extra-dimensional collapse, it will be hidden within an event horizon.
To give some insight into the mechanism by which this occurs, we also
give an explicit example of a collapsing spacetime and try to make
plausible its subsequent evolution into a black string.

Our argument relies on the cosmic censorship conjecture in four
space-time dimensions. This conjecture asserts, roughly, that all
singularities are hidden inside an event horizon rather than being
``naked,'' i.e., visible to distant observers; or, in other words,
that black holes are the generic final states of gravitational
collapse. Although it has not been proven, the cosmic censorship
conjecture is widely believed to be true for generic initial
conditions.\footnote{It is possible to construct non-singular
initial data for which the subsequent evolution contains a naked
singularity; however, analytic and numerical
studies~\cite{christodoulou94, choptuik} strongly suggest that such
initial data is in some sense non-generic.}

Ten years ago, Gregory and Laflamme~\cite{gregoryandlaflamme} showed
that black strings are, in fact, unstable to linear perturbations, at
least when the scale of the extra dimensions is large enough. If this
instability is a true, non-linear instability, the question then
arises as to what the final state will be. Gregory and Laflamme
suggested that the black string would ``fragment'' into a chain of
black holes, although, since this would require the event horizon to
bifurcate (a process that is forbidden if five-dimensional cosmic
censorship holds), a naked singularity would result. Thus there is
something of a puzzle as to what the final state actually is: if one
imposes the symmetry constraint that we do, the final state appears to
be a black string; if one does not, then a naked singularity appears
to be possible. It has also been suggested~\cite{horowitz} that the
instability will not lead to a bifurcation of the event horizon and
that, instead, the spacetime evolves to a stable solution that does
not have translational symmetry in the extra dimension.

The outline of this paper is as follows: In
\refsec~\ref{chap:conformal} we describe the cosmic censorship
conjecture and the conditions under which it is believed to hold. We
then rewrite Einstein's equation for the five-dimensional spacetime as
a four-dimensional theory with an effective matter content and show
that this effective matter content does indeed satisfy the conditions
of the cosmic censorship conjecture.  In
\refsec~\ref{chap:collapsemodel} we show how it is in principle
possible to construct initial data that is guaranteed to form a
singularity and then give, explicitly, a class of such initial
data. By considering a plausible scenario for the evolution of this
data, we illustrate how the black string likely arises.

\section{A General Argument from the Cosmic Censorship Conjecture}
\label{chap:conformal}

In the Introduction we gave a simple example of a spacetime possessing
an extra dimension in which the extra dimension collapses to zero size
everywhere on a spacelike surface and the worldline of every observer
ends on the singularity in finite proper time. In this \refsec\ we
argue that such a catastrophic fate will not befall more realistic
examples and that even a naked singularity will not occur, provided
that the extra dimension is homogeneous.

In order to proceed, we shall make the simplifying assumption that the
spacetime is the product of a four-dimensional manifold, $M$, with
$S^1$ (though it makes no difference to our argument if the extra
dimension has the topology of~$\R$) and that the metric of the full
spacetime, $\G_{AB}$ can be written in the form
\begin{equation}\label{eq:fullmetric}
ds^2 = \G_{AB}\,dx^A dx^B = \g_{ab}(x)dx^adx^b + e^{2\sqrt{2/3}\,\beta(x)}\,d\z^2,
\end{equation}
where the $x^a$ are coordinates in the ``ordinary,'' four-dimensional,
spacetime, $\z$ the coordinate in the extra dimension, and we shall
use uppercase Roman letters to denote indices in the full,
five-dimensional spacetime but lowercase Roman letters for indices in
the four-dimensional spacetime. 

The four-dimensional metric $\g_{ab}$, and the scale factor
$\beta(x)$, do not depend upon~$\z$ but are otherwise completely
general. That is, we consider only spacetimes in which the extra
dimension is homogeneous. This form of the metric is typical of many
models considered in the literature\footnote{There are exceptions,
notably those with a non-factorizable, ``warped''
metric~\cite{rs,rs1}.} and is similar to the original Kaluza-Klein
ansatz except that we disallow off-diagonal terms in the metric.

Einstein's equation in the full spacetime, in geometric units (where
$G=c=1$), is
\begin{equation}\label{eq:fulleinstein}
\Eins_{AB}=8\pi\Stress_{AB},
\end{equation}
where $\Eins_{AB}$ is the five-dimensional Einstein tensor, and
$\Stress_{AB}$ the five-dimensional stress-energy tensor. Note that to
be consistent with the form of $\G_{AB}$ given above we must impose
the condition $\Stress_{a\z}=0$ on the stress-energy. 

Our approach will be to show that this equation may be rewritten as
the equations describing four-dimensional relativity with the addition
of a scalar field, and hence to argue that the four-dimensional cosmic
censorship conjecture precludes the existence of either a naked
singularity or a spacetime-destroying one. This ``dimensional
reduction'' is usually carried out in a Lagrangian formulation (see,
for example, the survey article by Overduin and Wesson~\cite{kksurvey}
and references therein) but we shall instead directly rewrite
Einstein's equation to arrive at a four-dimensional theory with some
effective stress-energy tensor. Rewriting Einstein's equation in this
way has the benefit that it is more straightforward to determine the
effective stress-energy tensor---particularly when the matter content
does not have a Lagrangian formulation---and, furthermore, one can be
sure of obtaining all the equations of motion.\footnote{If one
substitutes a metric ansatz (such as eq.~(\ref{eq:fullmetric})) into
an action, subsequent variation of the action will not necessarily
give rise to all the equations of motion.}

\subsection{Dimensional Reduction}\label{sec:dimreduct}

From our metric ansatz, eq.~(\ref{eq:fullmetric}), we can rewrite the
five-dimensional tensors appearing in the theory in terms of their
four-dimensional counterparts. We find,
\begin{equation}
\begin{split}
\Ricci_{ab} &= \ricci_{ab}[\g_{ab}] - \frac{2}{3}D_a\beta D_b\beta 
		- \sqrt{\frac{2}{3}}D_a D_b \beta, \\
\Ricci_{\z\z} &=
		-e^{2\sqrt{2/3}\,\beta}
		\Bigl(\frac{2}{3}D^a\beta D_a\beta 
			+\sqrt{\frac{2}{3}}D^aD_a\beta\Bigr). 
\end{split}
\end{equation}
Here $\Ricci_{ab}$~is the five-dimensional Ricci tensor projected into
the four-di\-men\-sion\-al space and $\ricci_{ab}[\g_{ab}]$ is the Ricci
tensor associated with the four-dimensional part of the
metric,~$\g_{ab}$.  (The mixed-index terms, $\Ricci_{a\z}$, are zero.)
Finally, $D_a$ is the derivative operator associated
with~$\g_{ab}$. Using the above we can rewrite the Einstein tensor:
\begin{equation}\label{eq:reducedeinstein}
\begin{split}
\Eins_{ab} &= \Ricci_{ab}-\frac{1}{2}\Ricci\G_{ab} \\
	   &= \eins_{ab} - \Bigl[\frac{2}{3}D^a\beta D_a\beta 
			+ \sqrt{\frac{2}{3}}D^aD_a\beta\Bigr] \\
	&\qquad\qquad	+ \Bigl[\frac{2}{3}D^a\beta D_a\beta 
			+ \sqrt{\frac{2}{3}}D^aD_a\beta\Bigr]\g_{ab}, \\
\Eins_{\z\z} &= -\frac{1}{2}\ricci e^{2\sqrt{2/3}\,\beta}.
\end{split}
\end{equation}
where $\ricci=\ricci_{ab}\g^{ab}$
and~$\eins_{ab}=\ricci_{ab}-\frac{1}{2}\ricci\g_{ab}$.

One could at this point equate the right hand side of the first
equation above to the four-dimensional part of the stress-energy
tensor and consider the expression involving $\beta$ as part of an
effective stress-energy. However, this expression is not recognizable
as the stress-energy of, say, a scalar field. To rewrite the equation
so that the stress-energy is recognizable, we make the conformal
transformation
\begin{equation}\label{eq:conformal}
\g_{ab} = e^{-\sqrt{2/3}\,\beta}\tg_{ab}.
\end{equation}
The Ricci tensor and scalar then become
\begin{equation}\label{eq:conformalricci}
\begin{split}
\ricci_{ab} &= \tricci_{ab}[\tg_{ab}] 
		+\frac{1}{3}\tD_a\beta\tD_b\beta 
			+ \sqrt{\frac{2}{3}}\tD_a\tD_b\beta \\
	&\qquad	+\frac{1}{2}\Bigl[\sqrt{\frac{2}{3}}\tD^c\tD_c\beta
			-\frac{2}{3}\tD^c\beta\tD_c\beta\Bigr]\tg_{ab},
		\\
\ricci &= e^{\sqrt{2/3}\,\beta}\Bigl[\tricci + 3\sqrt{\frac{2}{3}}\tD^c\tD_c\beta 
		- \tD^c\beta\tD_c\beta\Bigr], \\
\end{split}
\end{equation}
where now $\tD_a$ is the derivative operator associated
with~$\tg_{ab}$ and indices are raised and lowered
with~$\tg_{ab}$. Finally, we substitute this expression for
$\ricci_{ab}$ into eq.~(\ref{eq:reducedeinstein}) and also replace
$D_a$ by $\tD_a$ there, to obtain
\begin{equation}
\begin{split}
\Eins_{ab} &= \teins_{ab} 
		- \tD_a\beta\tD_b\beta 
		+ \frac{1}{2}(\tD^c\beta\tD_c\beta)\tg_{ab}, \\
\Eins_{\z\z} &= -\frac{1}{2}e^{3\sqrt{2/3}\,\beta}
			\Bigl[\tricci + 3\sqrt{\frac{2}{3}}\tD^c\tD_c\beta 
		- \tD^c\beta\tD_c\beta\Bigr].
\end{split}
\end{equation}
Thus, from Einstein's equation in the full spacetime,
eq.~(\ref{eq:fulleinstein}), we have,
\begin{equation}
\begin{split}
\teins_{ab} &= 8\pi\Stress_{ab} + 
	 	\tD_a\beta\tD_b\beta 
		- \frac{1}{2}(\tD^c\beta\tD_c\beta)\tg_{ab}, \\
\tD^a\tD_a\beta &= -\sqrt{\frac{2}{3}}e^{-3\sqrt{2/3}\,\beta}
				\Stress_{\z\z}
		+\frac{1}{\sqrt{6}}\tg^{ab}\Stress_{ab},
\end{split}
\end{equation}
One may interpret this as the theory of General Relativity in four
dimensions, with matter content described by $\Stress_{ab}$, plus a
massless scalar field,~$\beta$, coupled to~$\Stress_{ab}$
and~$\Stress_{\z\z}$.

We now discuss the cosmic censorship conjecture.

\subsection{The Cosmic Censorship Conjecture}

It is widely believed that in a four-dimensional spacetime arising
from reasonable initial data, with reasonable matter content, no
singularities will be visible to distant observers; that is, all
singularities will be hidden within black holes. (See, e.g.,
Wald~\cite{waldsurvey} for a survey of past and recent results.) Here
we recall the precise statement of this conjecture by giving a meaning
to the notion of ``reasonable'' initial data, ``reasonable'' matter,
and ``distant observers,'' and hence argue that the singularity formed
by a collapsing extra dimension will likewise be hidden, given the
results of section~\ref{sec:dimreduct}.

We first say what is meant by a distant observer. The intuitive
meaning is an observer located ``far away, in the future'' where the
spacetime ``looks like'' flat spacetime. The precise meaning for these
terms is given by the notion of asymptotic flatness at future null
infinity. Roughly speaking, future null infinity, $\scri^+$, is the
``endpoint'' of null geodesics that propagate out to large
distances. (The details of this construction, which are not important
here, can be found in advanced textbooks on general
relativity~\cite[Chapter 11]{waldbook}.) If the spacetime is
asymptotically flat at future null infinity then it ``looks like''
flat spacetime at sufficiently large distances and late times;
$\scri^+$ then represents ``far away in the future.'' The notion that
a distant observer will be able to avoid running into a singularity is
then captured by the precise statement that future null infinity is
complete. Furthermore, if no past-directed causal curve from $\scri^+$
terminates at a singularity, then distant observers will not be able
to see the singularity.

Next we explain what sort of initial data we allow. Clearly no version
of the cosmic censorship conjecture will hold without some restriction
on the initial data: for example, the spacetime given in the
Introduction \emph{does} produce a spacetime-destroying
singularity. On the other hand, if one lives in a spacetime that is
not, initially, collapsing everywhere, one cannot create such initial
collapse because the collapse is not confined to some compact
region. We thus wish to require that at large distances the initial
data approaches flat space. It turns out that a notion of asymptotic
flatness may be defined for initial data sets, analogous to asymptotic
flatness at future null infinity for spacetimes, and we will allow
only asymptotically flat initial data.

Finally, for the purposes of the conjecture, the matter content must
be ``well-behaved'' in the following sense: 
\begin{enumerate}
\item\label{item:initialvalue} 
The coupled Einstein-matter equations have a well-posed initial value
formulation;
\item\label{item:energy} 
The matter satisfies the dominant energy condition so that observers
do not see negative energy densities or ``superluminal'' energy flow;
and
\item\label{item:nonsingular}
The matter is not of such a nature as to produce singularities in a
fixed, non-singular, background spacetime, uncoupled from Einstein's
equation.
\end{enumerate}

We now state one version of the cosmic censorship
conjecture.\footnote{There are actually two closely related
conjectures: the one we give is known as the weak cosmic censorship
conjecture. The strong cosmic censorship conjecture says, roughly,
that no one ever sees a singularity---not even one who falls into a
black hole---unless he runs into it.}

\emph{Weak Cosmic Censorship Conjecture:} 
Consider asymptotically flat initial data for Einstein's equation with
suitable matter, in the sense given above. Then, generically, the
maximal Cauchy evolution of this data is a spacetime that is
asymptotically flat at future null infinity, with complete~$\scri^+$.

\subsection{Application to Extra-Dimensional Spacetimes}

If the extra dimension collapses in the evolution of a
five-dimensional spacetime whose metric is of the
form~(\ref{eq:fullmetric}) then a singularity will be produced. In the
conformally transformed, four-dimensional theory, this singularity
appears as a divergence of the scalar field and, in particular, a
divergence of the stress-energy of the scalar field. Thus, there will
also be a space-time singularity in the four-dimensional
theory. However, we are now in a position to argue that this
singularity will be contained within a black hole.

Thus, consider a five-dimensional spacetime for which the
five-dimensional matter content satisfies conditions
\ref{item:initialvalue}--\ref{item:nonsingular} above and such that the
initial data for the equivalent four-dimensional spacetime is
asymptotically flat; then, assuming that the cosmic censorship
conjecture is true, we claim that the singularity will be contained
within a black hole (a black string in the five dimensional theory).

To see that this is true, let there be given initial data for the
five-dimensional spacetime and hence, by the equivalence described in
section~\ref{sec:dimreduct}, we obtain initial data for a
four-dimensional spacetime with matter content that includes a scalar
field. Now to show that the cosmic censorship conjecture applies to
the four-dimensional initial data we must show that the
four-dimensional matter content is well-behaved in the sense above. It
is well-known that a massless scalar field is well-behaved. Since, by
assumption, the five-dimensional equations have a well-posed initial
value formulation it is clear that the four-dimensional equations will
also, for they are just a rewriting of the five-dimensional
equations.

Likewise, note that the evolution of this matter from non-singular
initial data in a fixed background with fixed~$\beta$ is equivalent to
that obtained by fixing the five-dimensional background spacetime and
thus will not produce a singularity. To satisfy
condition~\ref{item:nonsingular} above we should actually fix only the
four-dimensional spacetime whilst allowing both the matter and~$\beta$
to evolve; but this is not equivalent, in the five-dimensional view,
to fixing the five-dimensional background spacetime. However, noting
that $\beta$ is, on its own, well-behaved, we would expect that, were
we also to allow $\beta$ to evolve, a singularity would not
arise. Thus, it appears highly plausible that
condition~\ref{item:nonsingular} does hold for the effective,
four-dimensional matter content.

It remains only to check that the four-dimensional stress-energy
satisfies the dominant energy condition. To this end, let $\xi^a$ be
any future-directed, timelike vector (future-directed and timelike
with respect to~$\tg_{ab}$). We must shown that the vector
$-\xi_aT_{bc}\tg^{ab}$ is future-directed timelike or null. But this
is true because $\xi^a$ is future-directed and timelike with respect
to $\g_{ab}$, and hence with respect to $\G_{AB}$, and, by assumption,
the dominant energy condition holds with respect to~$\G_{AB}$.

Thus, if the four-dimensional cosmic censorship conjecture holds, the
singularity formed in the four-dimensional spacetime with
metric~$\tg_{ab}$ will be contained within a black hole.

Now note that the projection of a curve in the five-dimensional
spacetime that is timelike (or causal) with respect to $\G_{AB}$ is a
curve in the four-dimensional spacetime that is timelike (or causal)
with respect to~$\tg_{ab}$. Thus, a reasonable definition of a
``distant observer'' in the five-dimensional spacetime would be one
whose world-line, when projected into the four-dimensional spacetime,
is the world-line of a distant observer there. Then, by the same
reasoning, if a distant observer in the five-dimensional spacetime
were able to see the singularity, the observer in the four-dimensional
spacetime obtained by projecting his worldline would be able to see
the singularity also. But we have already argued that the
four-dimensional distant observers do not see the singularity and we
therefore conclude that the singularity is not visible to distant
observers in the full spacetime, either.

\subsection{More than One Extra Dimension}

To conclude this \refsec, we comment briefly on an obvious
generalization of this model to more than one extra dimension. It
turns out that in this case the conclusion that the effective matter
content satisfies the dominant energy condition does not necessarily
hold.

Suppose that there are now $n$ extra dimensions. Previously we
required that the scale factor $e^{\beta(x)}$ did not depend on the
coordinate of the extra dimension. Likewise here, for simplicity, we
shall assume that the extra-dimensional manifold is a maximally
symmetric space whose metric depends on position only through a scale
factor that varies with position in the ``usual'' dimensions. The
metric then has the form
\begin{equation}
ds^2 = \g_{AB}dx^Adx^B+e^{\sqrt{\frac{2}{n(n+2)}}\,\beta}\gamma_{\mu\nu}dy^\mu dy^\nu,
\end{equation}
where the $y^\mu$ are the coordinates in the $n$ extra dimensions and
$\gamma_{\mu\nu}$ is the metric of a maximally symmetric manifold. The
action for General Relativity in this model may be dimensionally
reduced in precisely the same way as shown above for the $n=1$ case
producing again an effective four-dimensional theory containing a
scalar field. (As in the case of one extra dimension, this reduction
is typically done in a Lagrangian picture \cite{classicalstab, kolb90,
guenther97}.) After dimensional reducing the equations and making the
conformal transformation
$\g_{ab}=e^{-\sqrt{2n/(n+2)}\,\beta}\tilde{g}_{ab}$ the result for the
effective stress energy is
\begin{equation}
\begin{split}
8\pi\stress_{ab} &= 8\pi\SStress_{ab} +
			 \tD_a\beta\tD_b\beta \\
	& \quad	{}-\frac{1}{2}\bigl(\tD^c\beta\tD_c\beta - R_\gamma
			 e^{-\sqrt{\frac{2(n+2)}{n}}\,\beta}\bigr)\tg_{ab},
\end{split}
\end{equation}
where $R_\gamma$ is the Ricci scalar of~$\gamma_{ij}$ and
$\SStress_{ab}$ the four-dimensional projection of the
$(4\!+\!n)$-dimensional stress energy. In contrast to the case of one extra
dimension, the scalar field part of the stress-energy contains a
``potential term'' $V(\beta)=-R_\gamma
e^{-\sqrt{2(n+2)/n}\,\beta}$. The scalar field will satisfy the
dominant energy condition if this potential is positive (that is, if
the manifold of extra dimensions is negatively curved or flat) but
will not do so if the potential is negative; that is, when the
manifold of extra dimensions has positive curvature.\footnote{There
are some indications that the more interesting case is that when the
curvature is non-negative, for only then does there exist spatially
homogeneous, static solutions to Einstein's equation
\cite{classicalstab}.} Typically, in models of extra-dimensional
cosmology, a matter term is included in the model (for instance to
stabilize the extra dimensions). It may then be the case that the
overall effective potential is positive even though the extra
dimensions have positive curvature (this is so, for example, in the
``monopole'' model~\cite{stab, classicalstab}).

\section{A Concrete Model of Extra-Dimensional Collapse}

\label{chap:collapsemodel}
In \refsec~\ref{chap:conformal} we concluded that local
extra-dimensional collapse would not give rise to a
spacetime-destroying singularity. In the Introduction we gave an
example of a spacetime whose extra-dimensional collapse \emph{did}
destroy the universe but in that example the collapse was not
initially confined to some local region. In this \refsec\ we turn the
example of the Introduction into a more pertinent one by constructing
a class of initial data for spacetimes in which the collapse does
occur locally. We then consider this initial data from the
four-dimensional point of view. 

Under this interpretation the size of the extra dimension appears as a
scalar field, constant everywhere on the initial data surface but
having non-zero time derivative in the inner region, where, moreover,
its value becomes~$-\infty$ after finite proper time. Thus, in this
picture, the stress-energy in the inner region becomes infinite and
hence a singularity forms.

Recall that an outer marginally trapped surface is a spacelike,
two-di\-men\-sion\-al submanifold that is the boundary of a
three-dimensional closed region, such that the expansion of the family
of outgoing null geodesics normal to the surface is non-positive. The
four-dimensional censorship conjecture implies that any outer
marginally trapped surface will be contained within, or coincident
with, an event horizon. The event horizon of a stationary black hole,
for instance, is an outer marginally trapped surface. From the
arguments given in
\refsec~\ref{chap:conformal}, we would, therefore, expect that an
outer marginally trapped surface will either exist in the initial
data, or be formed sufficiently early to enclose any singularity.

In this section we show how the choice of initial data (within our
class of models) that ensures extra-dimensional collapse, also leads
one to the conclusion that an outer marginally trapped surface will
surround the singularity. In doing so, we gain some insight into the
``mechanism'' by which the conclusions of \refsec~\ref{chap:conformal}
are enforced.

The idea of our construction is to give initial data that, within some
compact region, ``looks like'' the collapsing initial data given in
the Introduction, but is then asymptotically flat outside this
region. Inside the region the spacetime does not ``know'' that the
rest of the spacetime has been changed, and the region will be made
large enough that collapse will be guaranteed to occur at some point
within it before information about that change propagates in. We now
make this idea precise.

Let $A$ be a compact subset of the $t=0$ hypersurface of the spacetime
described in the Introduction. (The set $A$ will be the ``region
within which collapse occurs.'') By the
\emph{future domain of dependence} of $A$ we mean the collection of
all points $p$ in the spacetime such that every past-directed,
inextendible,\footnote{For the definition of inextendible see
Wald~\cite[Chapter 8]{waldbook}. One can always find a curve from $p$
that does not intersect $A$ by taking one that does and letting it end
before it reaches~$A$. The technical restriction of inextendibility
prevents this kind of ``cheating.''} causal curve through $p$
intersects~$A$. We denote the future domain of dependence of~$A$ by
$D^+(A)$. The point of this definition is that properties of the
spacetime at some point $p\in D^+(A)$ (such as the metric and any
matter fields) depend only upon the initial data specified on~$A$
since an observer at $p$ cannot ``see'' any other part of the initial
data set. Furthermore, if there is given new initial data having some
region $A'$ within which the data is the same as that within $A$,
then, in the resulting spacetimes, the two regions $D^+(A)$ and
$D^+(A')$ will be isometric.

Hence, if the extra dimension collapses to zero size within $D^+(A)$
then collapse will also occur within $D^+(A')$, i.e., we are
guaranteed that the collapse will, in fact, occur in the spacetime
resulting from the new initial data.

\begin{figure}
\includegraphics{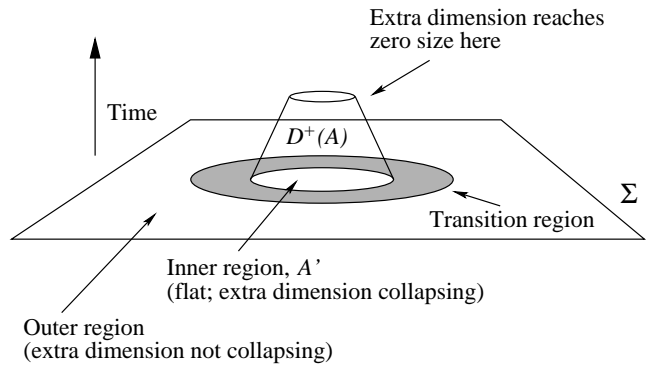}
\caption{A schematic diagram of the spacetime described in the
text. One ordinary spatial dimension and the extra dimension have been
suppressed. The initial data surface is labelled $\Sigma$. The initial
stress-energy is non-zero only in the shaded
region.\label{fig:initial}}
\end{figure}

In the next section, we describe initial data having this property. A
schematic diagram of the resulting spacetime is shown in
Figure~\ref{fig:initial}. The initial data surface is labelled
$\Sigma$; there is an inner region in which the space is flat but in
which, however, the extra dimension is collapsing uniformly (this is
the region equivalent to~$A$) where we have chosen the region large
enough that we can be sure that the extra dimension reaches zero size
within its future domain of dependence.

It might appear that the construction of such initial data would be
almost trivial: one sets the metric to be Minkowski out near infinity;
inside some region one sets the metric to be such that the extra
dimension is collapsing and then smoothly joins these two regions,
allowing the matter content of the spacetime to be determined by
Einstein's equation. However, this procedure would likely result in
``unreasonable'' matter, having negative energy densities or
superluminal energy flow. We shall insist that the matter content
satisfy the dominant energy condition (see section~\ref{sec:dominant})
(this is the same condition as that used in
\refsec~\ref{chap:conformal}).


The outline of the remainder of this \refsec\ is as follows: In
section~\ref{sec:initialdata} (and Appendix~\ref{chap:rounding}) we
describe initial data whose extra dimension will collapse to zero
size. In sections~\ref{sec:collapse} and \ref{sec:dominant} we write
down the conditions that the spacetime will collapse and that the
matter does satisfy the dominant energy condition and in
section~\ref{sec:trapped} we give the condition that an outer
marginally trapped surface be present in the initial data. Finally, in
section~\ref{sec:visible} we argue that, even in those cases where the
initial data does not contain such a surface, there is good reason to
believe that the future evolution will contain an event horizon that
will prevent signals from the singularity reaching infinity.
 

\subsection{Initial Data Guaranteed to Collapse}
\label{sec:initialdata}
\subsubsection{General Considerations}

In \refsec~\ref{chap:conformal} we considered spacetimes whose
manifold structure was the metric product of a four-dimensional
manifold, $M$, with a circle, such that the metric did not depend on
the circle coordinate. Here also we shall only consider spacetimes
with this property. Furthermore we suppose that the initial data, and
hence the resultant spacetime, is spherically symmetric.\footnote{By
spherically symmetric we mean that there exists an action of the group
$\mathrm{SO}(3)$ as an isometry on the spacetime whose orbits are
(spatial) 2-spheres.} We now introduce a convenient coordinate system
for such a spacetime.

Since the spacetime is spherically symmetric we introduce coordinates
$r$, $\theta$, and $\phi$ in the usual way (so that for any point
whose radial coordinate is~$r$, the area of the 2-sphere of spherical
symmetry containing that point is $4\pi r^2$). The coordinate in the
extra dimension we again denote by~$\z$.

Assuming that the constant-$r$ surfaces are timelike, the spacetime
can now be foliated by four-dimensional (``constant time'')
hypersurfaces $\Sigma_t$ such that $\nabla_a r$ is orthogonal
to~$\nabla_a t$. (That is, a surface invariant both under rotations
and translation in the extra dimension is a constant time surface if
the integral curves of $\nabla_a r$ lie within it.)

If the constant-$r$ surfaces become null (e.g., the $r=2M$ surface in
Schwarzschild) this construction is not valid since the normal now
lies in the surface and the $t$ and $r$~coordinates become
degenerate. Likewise this construction may also fail where
$\nabla_ar=0$ for then $r$ no longer necessarily identifies uniquely a
single constant-$r$ surface. However, in the region of the initial
data surface that we give later, neither of these problems arises.

It follows that in this coordinate system the metric may be written,
\begin{equation}\label{eq:fullmetric2}
\begin{split}
ds^2 &= \G_{AB}dx^A dx^B \\
	&= -f(r,t)^2\, dt^2 + g(r,t)^2 \,dr^2 +
		r^2\,d\Omega^2 + h(r,t)^2\,d\z^2. 
\end{split}
\end{equation}  
Here $d\Omega^2= d\theta^2 +\sin^2\theta\,d\phi^2$ is the spherical
part of the metric. In section~\ref{subsec:models} the functions
$f(r,t)$, $g(r,t)$, and $h(r,t)$ will be restricted further by a
choice of a class of initial data sets.

Our initial data surface will be $\Sigma=\Sigma_0$. The induced metric
on $\Sigma$ is clearly,
\begin{equation}
s_{AB}dx^A dx^B = g(r,t)^2\, dr^2 + r^2\, d\Omega^2
        + h(r,t)^2\,d\z^2.
\end{equation}
Now, letting $n^A=-f^{-1}(\partial/\partial t)^A$ be the field of
unit, timelike vectors orthogonal to $\Sigma$, so that
$s_{AB}=\G_{AB}+n_An_B$, we compute the extrinsic curvature of
$\Sigma$ from the usual formula,
\begin{equation}
K_{AB} = \nabla_A n_B.
\end{equation}
Some algebra gives
\begin{equation}\label{eq:K}
K_{AB} = -\frac{g\dot{g}}{f} (dr)_A(dr)_B 
	-\frac{h\dot{h}}{f}(d\z)_A(d\z)_B,
\end{equation}
where a dot denotes a derivative with respect to~$t$.

We shall take the matter to be dust, for simplicity.\footnote{It is
possible for the evolution of dust from non-singular initial data in a
fixed background to produce a singularity: dust does not satisfy the
third of the conditions on the matter content in the cosmic censorship
conjecture. However, our intent is not to illustrate a naked
singularity but the formation of an outer marginally trapped surface
and the failure of dust to satisfy property~\ref{item:nonsingular}
will not be relevant to our considerations.} Such a choice has the
advantage that the equation of state is trivial (the pressure is zero)
as is the equation of motion (the ``dust particles'' follow
geodesics). The matter stress-energy is then of the form
\begin{equation}
\Stress_{AB}=\sigma u_Au_B
\end{equation}
for some density $\sigma$ and four-velocity $u^A$; for consistency
with our metric ansatz, we must assume $u_\z=0$. (We write $\sigma$
for the energy density in the stress-energy of the dust to distinguish
it from the initial-data energy density.) 

The initial, five-dimensional, energy and current densities,
$\rho=T_{AB}n^An^B$ and $J_A=-h{}_A{}^CT_{CB}\xi^B$ are given by the
same expressions as in four dimensions (see, e.g., Wald \cite[chapter
10]{waldbook}):
\begin{align}
16\pi\rho &= \fourRicci + (K{}^A{}_A)^2 - K_{AB}K^{AB}, \\ 
-8\pi J_A &= D^B(K_{AB} -K{}^C{}_Ch_{AB}).
\end{align}
Here $D_A$ is the spatial derivative operator (i.e., the derivative
operator on $\Sigma$ associated with $s_{AB}$) and $\fourRicci$ is the
curvature scalar for $(\Sigma, s_{AB})$. Substituting in the equation
above the formula (\ref{eq:K}) for the extrinsic curvature gives,
after some work,
\begin{align}\label{eq:rho}
16\pi\rho &= \frac{2}{r^2}\frac{\partial}{\partial r}
	\left[r(1-g^{-2})\right] -
	\frac{2}{hgr^2}\frac{\partial}{\partial r}\Bigl(
	\frac{r^2h'}{g}\Bigr) + \frac{2\dot{g}\dot{h}}{f^2gh}, \\
8\pi J_r &= \frac{\dot{g}}{fghr^2}\frac{\partial}{\partial r}
		(r^2h) - \frac{1}{h}\frac{\partial}{\partial r}
			\Bigl(\frac{\dot{h}}{f}\Bigr)\label{eq:j},
\end{align}
where a prime denotes a derivative with respect to~$r$. (The radial
component of $J_A$ we have written $J_r$; all the other components are
zero.)

\subsubsection{A Particular Class of Models}\label{subsec:models}

We now construct initial data for a class of spacetimes whose extra
dimension is collapsing. For some examples in this class, the extra
dimension will be guaranteed to collapse and, by looking for outer
marginally trapped surfaces, we shall gain some insight into how
distant observers are shielded from the extra-dimensional collapse,

This initial data may be described as follows. The space is
spherically symmetric and is divided into three regions: An interior
region, $r<r_1$, in which the size of the extra dimension is
collapsing at a constant rate; a transition region, $r_1\leq r\leq
r_2$; and an outer region, $r>r_2$, that is the metric product of the
exterior of the Schwarzschild solution and the extra dimension. In the
transition region we shall choose the metric functions to interpolate
in a simple way between their values at $r_1$ and~$r_2$.

The particular forms of the metric functions are given in
Table~\ref{tab:initial}. As well as $r_1$ and $r_2$, three other
parameters determine the initial data: $h_0$,~the initial size of the
extra dimension (which is everywhere the same); $v_0$,~the initial
speed of the collapse of the extra dimension in the interior region;
and $M$, the mass per unit length.\footnote{That is, the exterior
region is the metric product of the Schwarzschild solution of mass~$M$
with the extra dimension: this is what is meant by ``mass per unit
length.''} For convenience, we give, instead of the metric function
$g(r)$, a function $q(r)$, where
\begin{equation}
g(r) = \Bigl(1-\frac{q(r)}{r}\Bigr)^{-1/2}.
\end{equation}
(When we do not write the $t$ argument for the metric functions, we
mean their initial values at $t=0$; thus $q(r)=q(r,t=0)$ and so
forth.)

\begin{table*}
\begin{ruledtabular}
\begin{tabular}{@{}rlll@{}}
At $t=0$ & $r<r_1$ & $r_1<r<r_2$ & $r>r_2$ \\ \hline
$q(r)$ & 0 & $\displaystyle 2M\frac{r-r_1}{r_2-r_1}$ & $2M$ \\
$f(r)$ & $(1-2M/r_2)^{1/2}$ & $(1-2M/r_2)^{1/2}$ & $(1-2M/r)^{1/2}$ \\
$h(r)$ & $h_0$ & $h_0$ & $h_0$ \\
$\dot{h}(r)$ & $-v_0$ & $\displaystyle -v_0\frac{r_2-r}{r_2-r_1}$
& 0 \\
\hline
$16\pi\rho(r)$ & 0 & $4M/[r^2(r_2-r_1)]$ & 0 \\
$8\pi j_r(r)$ & 0 & $\displaystyle
-\frac{v_0}{h_0}(1-2M/r_2)^{-1/2}(r_2-r_1)^{-1}$ & 0 \\ \hline
$8\pi(J^aJ_a)^{1/2} = 8\pi g^{-1}j_r$ &
0 & $\displaystyle -\frac{v_0}{h_0(r_2-r_1)}
	\biggl[1-\frac{2M}{r}\frac{r-r_1}{r_2-r_1}\biggr]^{1/2}
	\biggl(1-\frac{2M}{r_2}\biggr)^{-1/2}$ & 0 \\
\end{tabular}
\end{ruledtabular}
\caption{Initial data for the metric functions and matter content of the spacetime described in the text.\label{tab:initial}} 
\end{table*}

Note that the metric functions are continuous, though not smooth,
across the boundaries at $r=r_1$ and $r=r_2$. In
Appendix~\ref{chap:rounding} we show that the corners may be rounded
off so that the metric is smooth everywhere, without affecting the
conclusions.

In the next two sections we describe the conditions our initial data
is supposed to satisfy: that collapse of the extra dimension be
guaranteed to occur and that the matter content satisfy the dominant
energy condition. It turns out to be convenient to introduce a
dimensionless measure of how fast the collapse of the extra dimension
is occurring; namely,
\begin{equation}
\gamma \equiv v_0r_1g_0/(h_0f_0).
\end{equation}
Without loss of generality, one could also set $r_1=1$ (though we
shall not do so) so that our class of spacetimes is described by just
three parameters: $M$, $r_2$, and~$\gamma$.

\subsection{Collapse}\label{sec:collapse}
The collapse will be guaranteed if it occurs within $D^+(A)$. The
boundary of $D^+(A)$ is defined by null rays emitted from the edge of
the inner region, $r=r_1$, at coordinate time $t=0$, and these will
reach $r=0$ at coordinate time $t=r_1/f$. Thus if the extra dimension
reaches zero size at $r=0$ before this time it cannot be prevented;
since the collapse occurs at $t=h_0/v_0$ we must have
\begin{equation}
\frac{h_0}{v_0} < \frac{r_1}{f}, 
\end{equation} 
or equivalently, 
\begin{equation}\label{eq:collapsecond}
\gamma > 1.
\end{equation}
(The collapse of the extra dimension might still occur even if this
condition is not satisfied, it is just that it will not occur in
$D^+(A)$ and thus cannot be guaranteed.)

\subsection{Dominant Energy Condition}\label{sec:dominant}
Recall that the dominant energy condition requires that the
stress-energy, $T_{ab}$, be such that, for all future directed,
timelike vectors $t^a$, the vector $-T{}^a{}_bt^b$ is a
future-directed timelike or null vector. For dust, with
five-dimensional stress-energy $\Stress_{ab}=\sigma u^au^b$, this
condition is equivalent to requiring that $\sigma\geq 0$ and that
$u^a$ is timelike which, in turn, is equivalent to $\rho^2\geq
J_aJ^a$. When applied to our initial data the condition is
\begin{equation}\label{eq:energycond}
\frac{2M}{r^2} \geq \frac{\gamma}{r_1}\left[
	1-\Bigl(\frac{2M}{r_2-r_1}\Bigr)
	\Bigl(\frac{r-r_1}{r}\Bigr)\right]^{\frac{1}{2}}, 
\end{equation}
and this must be true for all $r$ in the range $r_1\leq r\leq
r_2$. 

\subsection{Trapped Surfaces}\label{sec:trapped}

Up to this point we have been working with the full, five-dimensional
spacetime, in part because it was easy to decide when collapse of the
extra dimension was inevitable. However, our arguments are based on
the four-dimensional cosmic censorship conjecture so, now, consider
what the initial data looks like in the dimensionally reduced,
conformally transformed picture described in
section~\ref{sec:dimreduct}. We now derive the condition that no outer
marginally trapped surfaces exist in the initial data.

In fact, it is sufficient to consider only surfaces of constant
$r$-coordinate for the following reason: If any outer marginally
trapped surfaces exist, consider the union of all the regions bounded
by such surfaces. The boundary of this region, which must be
spherically symmetric, is also an outer marginally trapped
surface~\cite[Chapter 12]{waldbook}. Thus, if any outer marginally
trapped surface exists, a spherically symmetric marginally trapped
surface exists.

On a constant-$r$ surface, the induced metric is
$\tilde{\omega}_{ab}dx^a dx^b=h(r)r^2 d\Omega^2$, where the factor of
$h(r)$ comes from the conformal transformation. The outgoing, future
directed, null vector field $\xi^a$ normal to the surface is
\begin{equation}
\xi^a = h^{-1/2}f^{-1}\bigl(\ddt\bigr)^a +h^{-1/2}g^{-1}\bigl(\ddr\bigr)^a.
\end{equation}
Hence the expansion, $\theta$, of the geodesics tangent to this vector
field is
\begin{equation}
\theta = \frac{1}{2}\tilde{\omega}^{ab}\lie_\xi \tilde{\omega}_{ab} 
	= h^{-1/2}\Bigl(g^{-1}\frac{2}{r}
	  + f^{-1}\frac{\dot{h}}{h}
	  + g^{-1}\frac{h'}{h}\Bigr).
\end{equation}
On substituting in our forms for $f$, $g$, and~$\dot{h}$, and
requiring $\theta>0$, we find that the condition that there are no
outer marginally trapped surfaces is:
\begin{equation}\label{eq:trappedcond}
\frac{2}{r}\biggl[1-
	\Bigl(\frac{2M}{r_2-r_1}\Bigr)
	\Bigl(\frac{r-r_1}{r}\Bigr)\biggr]^{1/2}
> \bigl(\frac{\gamma}{r_1}\bigr)\bigl(\frac{r_2-r}{r_2-r_1}\bigr),
\end{equation}
where the inequality must hold for all $r$ such that
$r_1\leq r\leq r_2$.  

\subsection{Visibility of the Singularity}\label{sec:visible}

We now consider the question: are there any values of the parameters
of our model for which conditions (\ref{eq:collapsecond}),
(\ref{eq:energycond}), and (\ref{eq:trappedcond}) hold? That is, is
there an example for which the collapse occurs, the dominant energy
condition is satisfied, and there are no outer marginally trapped
surfaces? If there is not, these examples will illustrate very clearly
how naked singularities are avoided in extra-dimensional collapse; if
there is, we shall consider whether such a surface is likely to form
around the singularity in the subsequent evolution of the spacetime.

Consider these conditions when $r=r_1$. We require $\gamma>1$ to
guarantee that the collapse occurs, whilst the condition that there be
no outer marginally trapped surfaces reduces to $\gamma<2$. Thus,
without even considering the rest of the spacetime, some of the
parameters of the model are already severely restricted by requiring
that the collapse not cause the formation of an outer marginally
trapped surface.

Next consider the conditions at $r=r_2$. We obtain $2M/r_2<1$ from the
condition that there be no outer marginally trapped surfaces, whereas
from the dominant energy condition one can obtain $2M/r_2 \gtrsim
0.6\gamma^2(r_2/r_1)^2$. Thus there are also severe restrictions on
the size of the transition region.

Given that one obtains these fairly restrictive conditions merely from
considering the points $r=r_1$ and $r=r_2$, one might imagine that one
could rule out all possible models by considering the conditions at
all values of $r$. However, it turns out that it \emph{is} possible to
choose parameters $\gamma$, $r_1$, $r_2$, and~$M$ such that the three
conditions are satisfied at \emph{all} values of~$r$.

Nonetheless, the conditions are quite restrictive. Fixing $r_1$, for
instance, it follows from the conditions above that there must be a
certain minimum amount of matter in the transition region and,
furthermore, the transition region cannot be too large. Consider,
also, the expression for the current density, $j_r$, given in
Table~\ref{tab:initial}: it is clear that $j_r$ is always negative,
which implies that the matter must be infalling. In other words, there
must be a certain amount of infalling matter contained in a region
that is not too large.  

To get some idea of how plausible it is that the singularity will be
hidden, we now make a very crude estimate of the time at which an
outer marginally trapped surface will form, and show that, according
to this estimate, the singularity occurs later than and inside an
outer marginally trapped surface.
\begin{figure}
\includegraphics{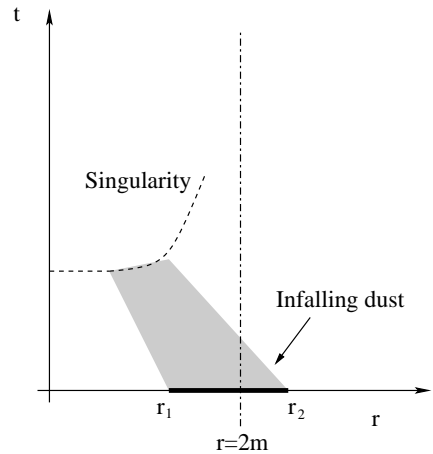}
\caption{The future evolution of the initial data described in the
text. The shaded area is the region containing matter. It is argued in
the text that the singularity will occur later than the formation of
an outer marginally trapped surface and inside
it.\label{fig:collapse}}
\end{figure}

In what follows we work in the full, five-dimensional spacetime where
it is easier to see when collapse of the extra dimension occurs.  A
schematic diagram of the spacetime is shown in
Figure~\ref{fig:collapse}. Referring to the metric,
eq.~(\ref{eq:fullmetric2}), and the initial forms of the metric
functions shown in Table~\ref{tab:initial}, one can see that if the
parameters describing the spacetime were chosen such that $r_2=2M$
then the surface $r=r_2$ would be an outer marginally trapped surface
and the exterior region would be that of a black string (for then the
metric in the exterior region would be the product of a black hole
spacetime of mass~$M$ with an extra dimension).

We shall therefore assume that an outer marginally trapped surface is
formed very roughly when the infalling dust passes $r=2M$. The initial
coordinate velocity of the outer surface of the dust is
$v_{\text{in}}=(f/g)(g^{-1}J_r/\rho)=
\gamma r_2^2\bigl(1-2M/r_2\bigr)/2M$ and so, very roughly, the dust
will reach $r=2M$ at a time $\tau_{\text{BH}}$, where
\begin{equation}
\tau_{\text{BH}} = \frac{r_2-2M}{v_{\text{in}}}f_0
		= \frac{2Mr_1}{\gamma r_2}f_0.
\end{equation}
(This is the proper time as measured by an observer at $r=2M$; the
factor of $f_0$ converts from coordinate time $t$ to proper time.)

On the other hand, for an observer at constant $r<r_1$, the
singularity will form at proper time $\tau_{\text{sing}}$, where
\begin{equation}
\tau_{\text{sing}} = \frac{h_0}{v_0}f_0 = \frac{r_1}{\gamma}.
\end{equation}
But now, noting that $2M/r_2<1$ and $f_0<1$, we have
\begin{equation}
\tau_{\text{BH}}<\tau_{\text{sing}}.
\end{equation}
It also follows from the dominant energy condition at $r=r_1$ that
$2M>r_1$, i.e., the inner region is within the radius at which we have
assumed an outer marginally trapped surface forms. (This is also
illustrated in Figure~\ref{fig:collapse}.)

\section{Conclusion}

For spacetimes that are the product of a four-dimensional spacetime
with an extra dimension, and for which the metric is independent of
the extra dimension, we have argued that collapse of the extra
dimension, though possible, will be hidden within a black string,
assuming that the four-dimensional cosmic censorship conjecture is
true.  We illustrated this conclusion with a class of examples in
which explicit initial data was given such that the extra-dimensional
collapse happened locally. For this class of examples it was clear
that ``trying to make the collapse happen sooner'' resulted either in
outer marginally trapped surfaces being present in the initial data
or, at any rate, a plausible collapse of the initial data to a black
string.

Presumably the resulting spacetime becomes nearly stationary at late
times. A well-known, black hole ``no-hair'' theorem
\cite{bekenstein72i, bekenstein72ii} asserts that the only stationary,
black-hole solutions to the Einstein--scalar field equations
necessarily have constant scalar field outside the black hole
horizon. Thus, if there were no matter content to the five-dimensional
spacetime (e.g., if it were all to fall in to the black hole or be
radiated away) this theorem would imply that the four-dimensional
spacetime resulting from extra-dimensional collapse has constant
scalar field; and this, in turn, implies that the five-dimensional
spacetime is a black string for which the size of the extra dimension
is constant. (If there is matter present the scalar field is
presumably not constant since it couples to the matter.) 

But this is just the type of spacetime considered by Gregory and
Laflamme and which, as mentioned in the Introduction, suffers from the
linear instability found by them. Thus, although we have assumed
four-dimensional cosmic censorship, the instability is evidence that
five-dimensional cosmic censorship does not hold.\footnote{If the
Gregory-Laflamme instability does lead to the violation of
five-dimensional cosmic censorship one cannot thereby immediately
obtain an example of a four-dimensional naked singularity by
dimensional reduction since the instability does not arise for an
homogeneous extra dimension.}

Nonetheless, there does not seem to be any good reason why cosmic
censorship should hold in four dimensions but not in five. If one
wanted to retain cosmic censorship in five dimensions then there seem
to be two possible ways of evading the dilemma. Perhaps the argument
that extra-dimensional collapse produces a black string fails for
inhomogeneous extra dimensions. Gregory and Laflamme have suggested
that the instability could set in before the black string forms,
giving rise, presumably, to one or more black holes, without horizon
bifurcation. It has also been argued~\cite{horowitz} that there is an
\emph{in}homogeneous, stable black string to which the homogeneous
black string will evolve.

On the other hand,
perhaps the black string scenario is the best place to look for an
explicit example of a (generic) naked singularity, albeit in five
dimensions. Such an example would presumably provide a great deal of
insight into the issue of cosmic censorship in four dimensions.

\begin{acknowledgments}
The work was supported in part by the \textsc{nsf} grant \textsc{phy}
00-90138 to the University of Chicago. The author is indebted to
Robert Wald for advice and encouragement.
\end{acknowledgments}

\appendix
\section{Rounding the Corners}\label{chap:rounding}

The metric functions described in \refsec~\ref{chap:collapsemodel} are
smooth in the interior region, $r<r_1$, in the transition region,
$r_1<r<r_2$, and in the exterior region $r>r_2$ but have discontinuous
first derivatives at $r=r_1$ and $r=r_2$. The purpose of this appendix
is to ``round off the corners,'' giving everywhere smooth functions
for which the existence of solutions to Einstein's equation is
guaranteed. Our smoothed functions will also have the property that
the smoothed metric will be equal to the original in the interior and
exterior regions except for small neighborhoods of $r_1$ and $r_2$,
which means that the extra-dimensional collapse is unaffected and the
exterior space is still Schwarzschild.

Some of the metric functions specified in
Table~\ref{tab:initial} are already smooth but the three that are not
are $q(r)$, $f(r)$, and $\dot{h}(r)$ (recall that $\dot{h}(r)$ is
specified directly as initial data; it is not calculated as the time
derivative of $h$). Now, the only form in which $\dot{h}(r)$ and
$f(r)$ enter into the dominant energy condition is as
$\dot{h}(r)/f(r)$; thus it is convenient to define
$s(r)=\dot{h}(r)/f(r)$ and to smooth $s(r)$ instead.\footnote{Once we
have obtained a smoothed $s(r)$ one may smooth $f(r)$ by any naive
method and then multiply it by the smoothed $s(r)$ to obtain a
smoothed~$\dot{h}(r)$.}

It is not hard to see that a piecewise smooth function may always be
smoothed out, in the sense that one can always find a smooth function
that is uniformly close to the given one. However, it is also clear
that the first derivative of a smooth approximation cannot be
uniformly close to the first derivative of the original function, for
the derivative of the original is discontinuous. Since the
stress-energy computed from the metric involves the first derivative
it is not at all obvious, and in general not true, that the
stress-energy computed from such a smoothed metric will satisfy the
dominant energy condition. (The collapse condition and the
no-trapped-surfaces condition will, however, be unaffected if the
region of rounding is made small enough.)

Our problem may therefore be stated as follows: Given metric functions
$q(r)$ and $s(r)$, satisfying appropriate conditions, find smooth
functions $\tilde{q}(r)$ and $\tilde{s}(r)$ such that the associated
stress-energy satisfies the dominant energy condition, which may be
written as
\begin{equation}\label{eq:varenergycond}
\frac{1}{r^2}\bigl(1-\frac{\tilde{q}}{r}\bigr)^{-1/2}
	\frac{d\tilde{q}}{dr} \geq \frac{d\tilde{s}}{dr}.
\end{equation}

The method we use to smooth the
metric functions is to convolve them with a smooth kernel. That is,
let $G(r)$ be a smooth, positive function with support in the region
$-1\leq r\leq1$ and with total integral unity. For any $\epsilon>0$,
define
\begin{equation}
G_\epsilon(r)=\frac{G(r/\epsilon)}{\epsilon},
\end{equation}
(noting that $G_\epsilon(r)$ also has total integral one) and set
\begin{equation}
\begin{split}
\tilde{q}_\epsilon(r) &= \int_{-\infty}^{\infty} q(r')G_\epsilon(r'-r)\,dr', \\
\tilde{s}_\epsilon(r) &= \int_{-\infty}^{\infty} s(r')G_\epsilon(r'-r)\,dr'.
\end{split}
\end{equation}

We claim that, for sufficiently small $\epsilon$, the functions
$\tilde{q}_\epsilon(r)$ and $\tilde{s}_\epsilon(r)$ satisfy the
dominant energy condition, eq.~(\ref{eq:varenergycond}), and,
furthermore, for $r<r_1-\epsilon$ and $r>r_2+\epsilon$ (the interior
and exterior regions respectively) we have $\tilde{q}(r)=q(r)$ and
$\tilde{s}(r)=s(r)$.\footnote{In the following, it may appear to be a
problem that the metric functions are defined only for $r\geq 0$,
whereas we write all formul\ae\ as if they were defined on the whole
real line. However, the functions to be smoothed are all constant for
$r<r_1$ so, if we choose $\epsilon<r_1$, the smoothed functions will
be unchanged near $r=0$.}  

To show this, we first define, for convenience,
\begin{equation}
F\bigl[q(r),r)\bigr] \equiv
\frac{1}{r^2}\bigl(1-\frac{q}{r}\bigr)^{-1/2},
\end{equation}
so that the dominant energy condition is
\begin{equation}
F\bigl[q(r),r\bigr]\frac{dq}{dr} \geq \frac{ds}{dr}.
\end{equation}
Now from the fact that $q(r)$ is uniformly continuous it follows that
$\tilde{q}_\epsilon(r)$ is uniformly approximated by $q(r)$, in the
sense that, given $\kappa>0$ there exists $\epsilon>0$ such that
$\lvert\tilde{q}_\epsilon(r)-q(r)\rvert<\kappa$ for all~$r$. From this and
the boundedness of $q(r)$, it follows that for any $\Delta>0$ there
exists $\epsilon>0$ such that for all $r'$ such that $\lvert r-r' \rvert < \epsilon$, 
\begin{equation}\label{eq:Festimate}
\bigl\lvert  F\bigl[\tilde{q}_\epsilon(r),r\bigr]- F\bigl[q(r'),r'\bigr] 
	\bigr\rvert < \Delta,
\end{equation}
and this bound is uniform in~$r$. We shall use this estimate in the
dominant energy condition.

For our metric functions, the inequality in the dominant energy
condition is saturated in the interior and exterior regions, where
both sides of the inequality are zero. When $r_1<r<r_2$, on the other
hand, the difference between the two sides is bounded away from
zero. That is, there exists $\delta>0$ such that, for $r_1<r<r_2$,
\begin{equation}\label{eq:vvarenergycond}
F\bigl[q(r),r\bigr]\frac{dq}{dr} \geq \frac{ds}{dr} +\delta.
\end{equation}
Choose $\Delta>0$ such that
\begin{equation}\label{eq:Delta}
\Delta\frac{dq}{dr}<\delta,
\end{equation}
(which is possible since $dq/dr$ is bounded) so that, using
(\ref{eq:Festimate}), we have,
\begin{equation}
\begin{split}
\biggl\lvert
 &\int F\bigl[q(r'),r'\bigr]\frac{dq(r')}{dr'}G_\epsilon(r'-r)\,dr' \\
& \quad {}-
\int F\bigl[\tilde{q}_\epsilon(r),r\bigr]\frac{dq(r')}{dr'}G_\epsilon(r'-r)\,dr'
\biggr\rvert \\ 
&\qquad< \int \Delta\frac{dq(r')}{dr'}G_\epsilon(r'-r)\,dr' \\
&\qquad< \delta \int G_\epsilon(r'-r)\,dr', 
\end{split}
\end{equation}
where the last line follows from eq.~(\ref{eq:Delta}).

Now convolve eq.~(\ref{eq:vvarenergycond}) with $G_\epsilon(r)$,
nothing that both sides are zero when $r<r_1$ and when $r>r_2$. Using
the estimate above, we find,
\begin{multline}\label{eq:useofstrictness}
F\bigl[\tilde{q}_\epsilon(r),r\bigr]
	\int_{r_1}^{r_2}\frac{dq(r')}{dr'}G_\epsilon(r'-r)\,dr' \\
+\delta \int_{r_1}^{r_2} G_\epsilon(r'-r)\,dr' \\
\geq  \int_{r_1}^{r_2} \frac{ds(r')}{dr'}G_\epsilon(r'-r)\,dr'
	+ \delta \int_{r_1}^{r_2} G_\epsilon(r'-r)\,dr'.
\end{multline}
The terms involving $\delta$ then cancel, and the integrands in the
remaining terms are zero outside the region of integration, so we may
take the limits of those integrals back to infinity. Thus,
\begin{multline}
F\bigl[\tilde{q}_\epsilon(r),r\bigr]
	\int\frac{dq(r')}{dr'}G_\epsilon(r'-r)\,dr' \\ 
\geq  \int \frac{ds(r')}{dr'}G_\epsilon(r'-r)\,dr'.
\end{multline}

Next, note a property of convolutions; namely, that
\begin{equation}
\begin{split}
\int\frac{dq(r')}{dr'}G_\epsilon(r'-r)\,dr' 
&= -\int q(r')\frac{d}{dr'}G_\epsilon(r'-r)\,dr' \\
&=  \int q(r')\frac{d}{dr}G_\epsilon(r'-r)\,dr' \\ 
&= \frac{d}{dr}\int q(r')G_\epsilon(r'-r)\,dr' \\
&= \frac{d\tilde{q}_\epsilon(r)}{dr},
\end{split}
\end{equation}
where, in the first line, we have integrated by parts.

Thus we find the desired result,
\begin{equation}
F\bigl[\tilde{q}_\epsilon(r),r\bigr] \frac{d\tilde{q}_\epsilon}{dr} 
\geq  \frac{d\tilde{s}_\epsilon}{dr}.
\end{equation}

Finally, we note that, for $r<r_1-\epsilon$ and $r>r_2+\epsilon$, both
$q(r)$ and $s(r)$ are constant and hence $\tilde{q}_\epsilon(r)=q(r)$
and $\tilde{s}_\epsilon(r)=s(r)$, as claimed.

\bibliography{geddes}

\end{document}